\documentclass[notitlepage,superscriptaddress, nofootinbib, prd, 10pt]{revtex4-1}

\usepackage{latexsym, amssymb, amsmath, graphics, graphicx, mathrsfs}
\usepackage{verbatim}
\usepackage[colorlinks=true,  linkcolor=black, citecolor=Green, urlcolor=black]{hyperref}
\usepackage[usenames,dvipsnames]{color}
\usepackage{multirow}
\usepackage{hepunits}

\newcommand{\be}{\begin{equation}}
\newcommand{\ee}{\end{equation}}
\newcommand{\eq}[1]{Eq.~(\ref{#1})}
\newcommand{\nl}{\nonumber \\}
\newcommand{\x}{\chi}
\newcommand{\qx}{q_\chi}
\newcommand{\w}{\omega}
\newcommand{\Ap}{A^\prime}
\newcommand{\mAp}{m_{A^\prime}}

\newcommand{\eps}{\epsilon}

\newcommand{\lap}{\nabla}
\newcommand{\grad}{\nabla}

\newcommand{\order}[1]{\mathcal{O}{(#1)}}

\newcommand{\E}{\boldsymbol{E}}
\newcommand{\B}{\boldsymbol{B}}
\newcommand{\vv}{{\bf v}}
\newcommand{\xv}{{\bf x}}

\newcommand{\zhat}{\hat{\boldsymbol{z}}}

\newcommand{\phihat}{\hat{\boldsymbol{\phi}}}
\newcommand{\jx}{\boldsymbol{j}_\x}

\newcommand\eea{\end{eqnarray}}
\newcommand\bea{\begin{eqnarray}}

\graphicspath{{figs/}}

\begin{document}

\title{Searching for Millicharged Particles with Superconducting Radio-Frequency Cavities}

\author{Asher Berlin}
\affiliation{Center for Cosmology and Particle Physics, Department of Physics, New York University, New York, NY 10003, USA}

\author{Anson Hook}
\affiliation{Maryland Center for Fundamental Physics, University of Maryland, College Park, MD 20742, USA}

\begin{abstract}
We demonstrate that superconducting radio-frequency cavities can be used to create and detect millicharged particles and are capable of extending the reach to couplings several orders of magnitude beyond other laboratory based constraints.
Millicharged particles are Schwinger pair-produced in driven cavities and quickly accelerated out of the cavity by the large electric fields.  The electric current generated by these particles is detected by a receiver cavity.
A light-shining-through-walls experiment may only need to reanalyze future data to provide new constraints on millicharged particles. 
\end{abstract}
\maketitle


\section{Introduction}
\label{sec:intro}

One of the most natural extensions of the Standard Model (SM) is to include a new charged particle with charge and mass different from the electron.  Particles of this type have been discovered before (for example, quarks) and may be discovered again.  If a new charged particle has an electric charge much smaller than the electron, then it is called a millicharged particle (mCP).  Due to their minimalistic properties, they often appear in theoretical extensions of the SM~\cite{Holdom:1985ag, Dienes:1996zr, Abel:2003ue, Batell:2005wa, Aldazabal:2000sa, Abel:2004rp,Abel:2008ai,Gherghetta:2019coi} and have often been used to explain various experimental anomalies~\cite{Zavattini:2005tm, Adriani:2008zr, Chang:2008aa, Barkana:2018lgd,Berlin:2018sjs,Barkana:2018cct,Liu:2019knx}. A natural way for such effective interactions to arise is through the kinetic mixing of a new light hidden sector dark photon, $\Ap$, with the SM photon,
\be
\mathscr{L} \supset \frac{\eps}{2} \, F_{\mu \nu} \, F^{\prime \mu \nu} + \frac{1}{2} \, \mAp^2 \, A_\mu^{\prime 2}
~,
\ee
where $\eps$ is a small dimensionless parameter that controls the strength of kinetic mixing and $\mAp$ is the dark photon mass~\cite{Holdom:1985ag}. On length-scales much smaller than $\mAp^{-1}$, the dark photon generates an effective millicharge under standard electromagnetism for particles $\x$ that are directly charged under the $\Ap$, of the form $\qx \simeq \eps \, e^\prime / e$
where $e^\prime$ is the $\Ap$ gauge coupling. From the perspective of such models, a small millicharge is a consequence of a small kinetic mixing parameter and/or hidden sector gauge coupling~\cite{Gherghetta:2019coi}.

There are many ways in which to produce mCPs.  Perhaps one of the most interesting production mechanisms is Schwinger pair-production~\cite{Schwinger:1951nm}.
In the presence of a large electric field, a particle and antiparticle can spontaneously appear.
If the electric field is larger than a critical value, then such particle-production is unsuppressed.  This critical electric field is 
\be
\label{eq:Ecr}
E_\text{cr} = \frac{m_\x^2}{e \qx} \sim 50 \ \text{MV} \ \text{m}^{-1} \times \left( \frac{m_\x}{\text{meV}} \right)^2 \left( \frac{q_\x}{10^{-7}} \right)^{-1}
~,
\ee
where $m_\x$ is the mass of the new particle and $\qx$ is its charge measured in units of the electron charge.  The largest laboratory based electric fields are currently many of orders of magnitude too small to produce any of the known particles at any appreciable rate.~\footnote{Producing electrons would require electric fields larger than $10^{12} \text{ MV} \text{ m}^{-1}$.}  However, if mCPs exist, then they might be produced by these large electric fields, and it behooves us to look for them.

\begin{figure}[t]
\begin{center}
\includegraphics[width=0.45\textwidth]{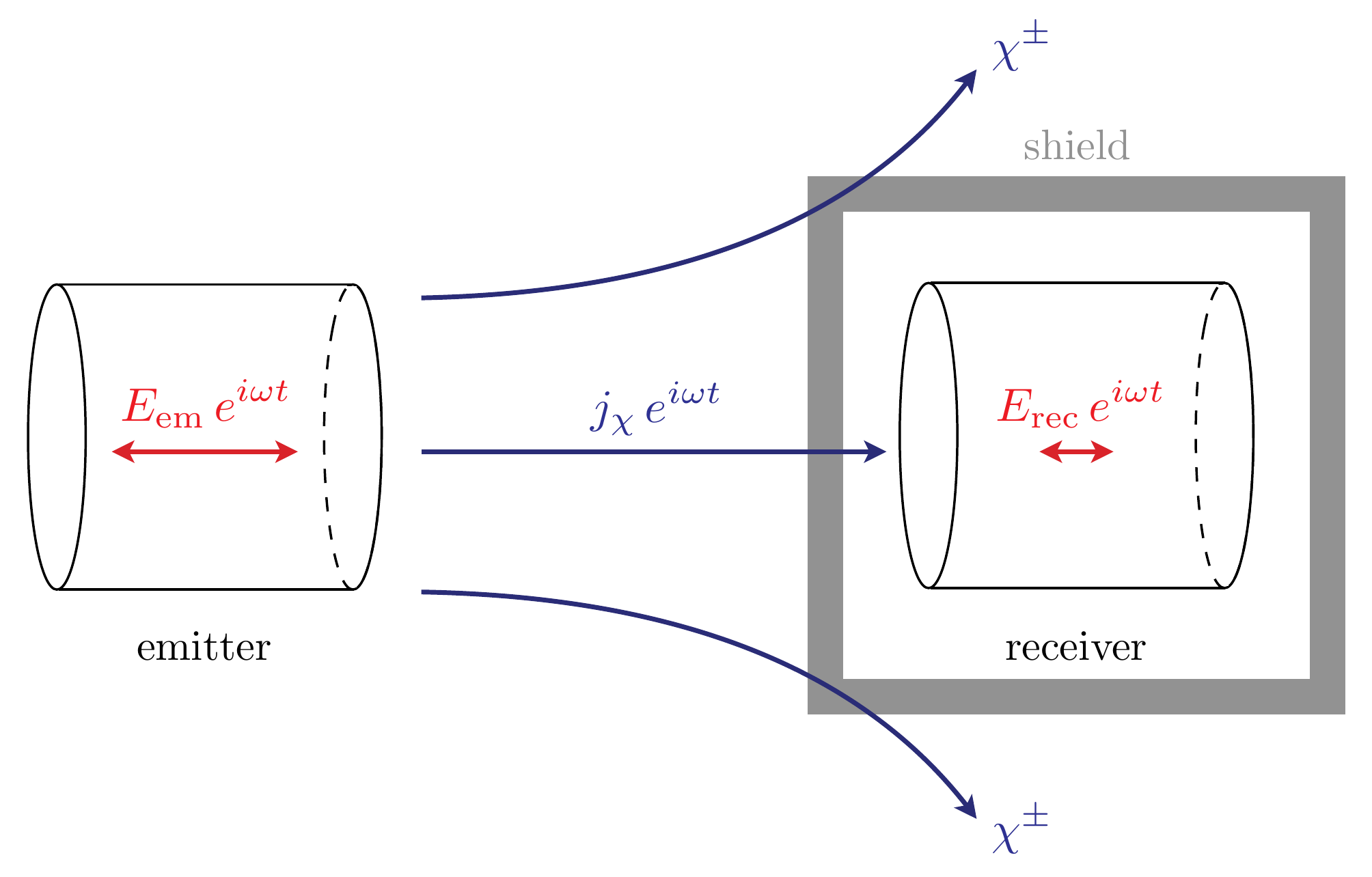}
\caption{A cartoon picture of our setup.  The large electric fields in the emitter cavity produce millicharged particles via Schwinger pair-production.  The electric field of the cavity is arranged such that the particles are accelerated towards the shielded receiver cavity, where the current of millicharged particles excites the resonant modes to detectable levels.}
\label{fig:cartoon}
\end{center}
\end{figure}

Using resonant cavities to search for new particles via Schwinger pair-production was first proposed in Refs.~\cite{Gies:2006hv,Jaeckel:2007hi}.
In this article, we propose using superconducting radio-frequency (SRF) cavities both to produce and detect mCPs.  The extreme environment of SRF cavities (with characteristic field strengths of $\sim 50 \ \text{MV} \text{ m}^{-1}$) makes them ideal for searching for new particles~\cite{Graham:2014sha,fnalth,fnalex,Bogorad:2019pbu,Janish:2019dpr,Berlin:2019ahk,Lasenby:2019prg}.  
We focus on a setup where a driven ``emitter" cavity operates in a mode where its electric field points towards a shielded ``receiver" cavity.  Millicharged particles are produced in the large electric field of the emitter cavity and are quickly accelerated out of the emitter cavity and towards the receiver cavity, easily penetrating an electromagnetic shield due their tiny electric charge.
The oscillating electric field of the driven emitter cavity imprints a characteristic frequency onto the produced current of mCPs.
The receiver cavity is tuned to have the same frequency, as to be resonantly sensitive to the oscillating mCP current.  The oscillating mCP current can
ring up the resonant modes of the receiver cavity to observable levels, constituting a discovery of mCPs.  A picture of the setup and its projected sensitivity are shown in Figs.~\ref{fig:cartoon} and \ref{fig:reach}, respectively.

A very interesting aspect of this setup is that it is nearly identical to a typical light-shining-through-walls (LSW) experiment, such as the one currently being constructed at Fermi National Accelerator Laboratory (FNAL) in order to search for ultralight hidden photons~\cite{Graham:2014sha,fnalth,fnalex}.
More generally, most LSW experiments can also be reinterpreted as an mCP search, regardless of whether or not an SRF cavity is utilized.  In this work, we show that near-future searches of this type are many orders of magnitude more sensitive to mCPs than other current laboratory based searches.

\section{Production of Millicharged Particles in Cavities}
\label{sec:schwinger}

Schwinger pair-production is the spontaneous appearance of a particle and antiparticle in the presence of a large electric field.  For particles produced at rest, this whole process conserves energy if the binding energy experienced by the particle-antiparticle dipole in the exterior electric field balances the rest mass energy,
\be
\label{eq:SchParam1}
e \qx \, d \, E \sim m_\x
~,
\ee
where $E$ is the external electric field and $d$ is the distance between the particle-antiparticle pair.
In quantum mechanics, everything that is allowed to happen can happen, but if there exists a large hierarchy in length-scales, the probability for such events to occur is exponentially suppressed. The length-scale associated 
with the virtual mCP pair is the Compton wavelength, $d_C \sim 1/m_\x$.
It is thus expected that Schwinger pair-production is exponentially suppressed if $d \gtrsim d_C$ and unsuppressed if $d \lesssim d_C$. This statement along with \eq{eq:SchParam1} can also be interpreted as demanding that the work performed by the electric field on the virtual mCP pair is sufficient to put the particles on-shell. From Eqs.~(\ref{eq:Ecr}) and (\ref{eq:SchParam1}), unsuppressed production ($d \lesssim d_C$) is equivalent to demanding $E \gtrsim E_\text{cr}$. This intuition is reflected in the expression for the probability of pair-creating particles per unit time and unit volume,
\begin{align}
\label{eq:schwinger}
P_\x &= \frac{dN_\chi}{dt ~ dV} \simeq \frac{c_\chi}{(2 \pi)^3} ~ (e q_\chi E)^2 ~ e^{- (\pi m_\x^2)/(e q_\chi E)}  \propto e^{-\pi d / d_C} \propto e^{-\pi E_\text{cr} / E}
\, ,
\end{align}
where $N_\x$ is the number of particle pairs and $c_\x = 1$ ($1/2$) for fermionic (scalar) mCPs~\cite{Schwinger:1951nm}. While this equation can only be rigorously defined in the limit when the exponential suppression is large, we will assume that it continues to hold even when the exponential suppression is not present.  This assumption can be shown to hold explicitly in the case of an electric field in a periodic box~\cite{Cohen:2008wz}.

For particles that are produced relativistically, \eq{eq:SchParam1} is modified to
\be
\label{eq:SchParam2}
e q_\x \, d_B \, E \sim p_\x
~,
\ee
where $p_\x$ is the mCP momentum and $d_B \sim 2 \pi \, p_\x^{-1}$ is its de Broglie wavelength. Solving \eq{eq:SchParam2} for $p_\x$ shows that the typical momentum of pair-produced particles is parametrically
\be
\label{eq:SchParam3}
p_\x \sim \sqrt{2 \pi \, e \qx E}
~.
\ee
Hence, $E \gg E_\text{cr}$ implies that $p_\x \gg m_\x$, i.e., electric fields much greater than the critical value lead to the production of relativistic particles~\cite{Cohen:2008wz}.

\begin{figure}[t]
\begin{center}
\includegraphics[width=0.7\textwidth]{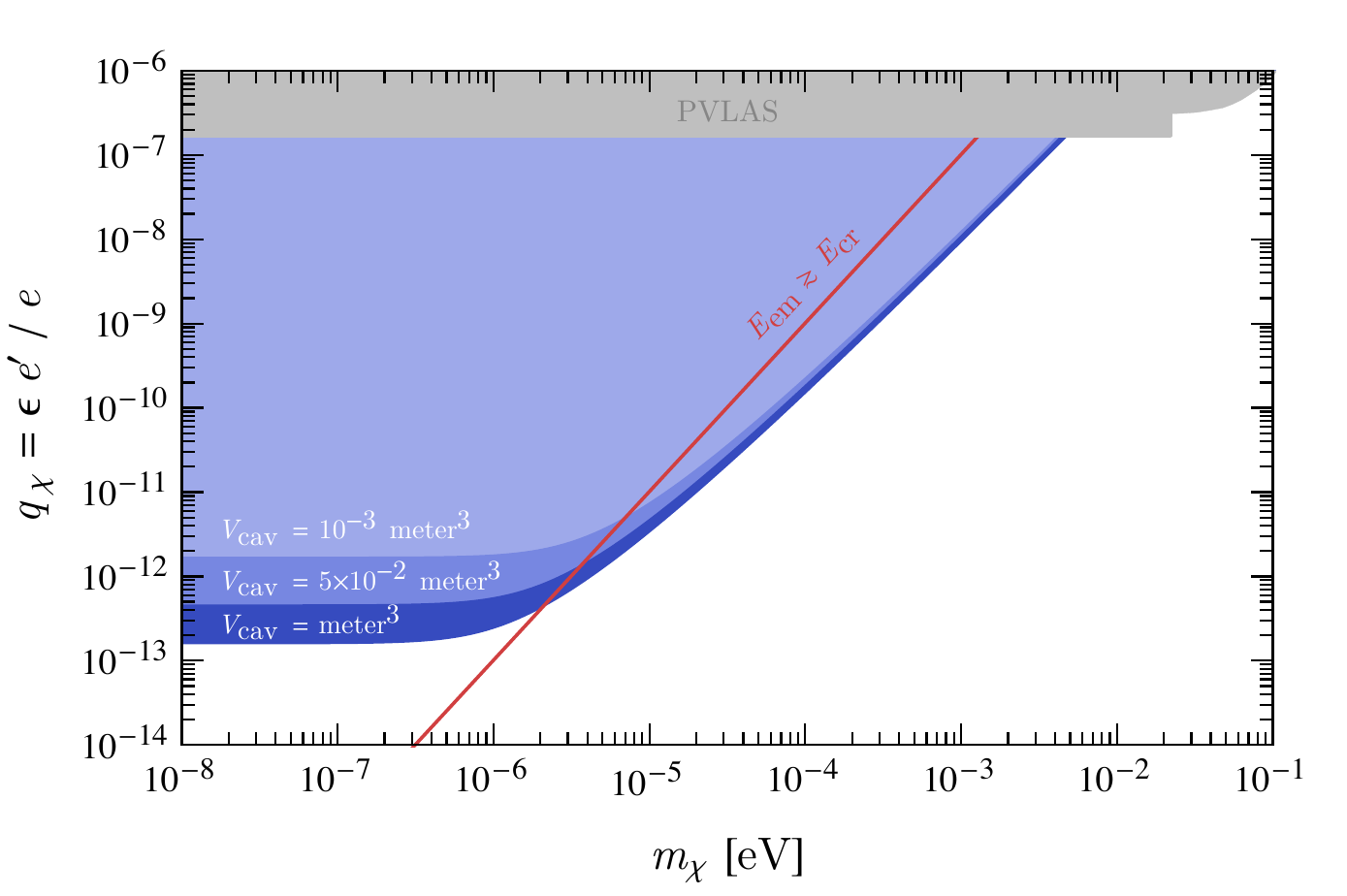}
\caption{The projected reach of future cavity experiments (shaded blue) to millicharged fermions for various volumes of the emitter/receiver cavities, $V_\text{cav}$ (the reach for millicharged scalars is weaker by a factor of $2^{1/3}$). In each case, we take the amplitude of the driven cavity field to be $E_\text{em} = 50 \text{ MV} \text{ m}^{-1}$, the quality factor to be $Q = 10^{12}$, the receiver cavity temperature to be $T = 10 \text{ mK}$, and the integration time to be $t_\text{int} = \text{year}$. We take the emitter and receiver cavities to both be cylinders of equal radius and length, such that the resonant frequency of the $\text{TM}_{010}$ mode is fixed to be $\w =  \alpha_{01} \left( \pi / V_\text{cav} \right)^{1/3}$, where $\alpha_{01}$ is the first zero of $J_0$.  The shaded gray region corresponds to the best existing laboratory bound from the PVLAS collaboration~\cite{Ahlers:2006iz,DellaValle:2014xoa}. 
Above the solid red line, the amplitude of the driven emitter cavity's electric field is larger than the critical field strength for Schwinger pair-production of millicharged particles. 
Not shown are astrophysical and cosmological limits derived from considerations of, e.g., stellar cooling, SN1987A, Big Bang nucleosynthesis, and the cosmic microwave background~\cite{Vogel:2013raa}. Models in which these constraints are mitigated are discussed in Appendix~\ref{sec:stellar}.}
\label{fig:reach}
\end{center}
\end{figure}

The production rate of \eq{eq:schwinger} applies to static and uniform electric fields of large spatial extent. However, in cavities, an oscillating electric field is confined to a finite interior region. Hence, in addition to $E \gtrsim E_\text{cr}$, efficient pair-production of mCPs requires that the characteristic length-scale associated with production is smaller than the typical length-scale over which the electric field varies by an $\order{1}$ fraction. For the lowest lying cavity modes discussed below, the latter is roughly $\w^{-1}$, where $\w \sim \GHz$ is the mode's resonant frequency. Therefore, for particles produced at rest (see \eq{eq:SchParam1}), pair-production is unsuppressed by spatial gradients of the cavity's electric field provided that the typical dipole length satisfies $d \lesssim \w^{-1}$, i.e.,
\be
\label{eq:adiabatic1}
\frac{\w \, m_\x}{e \qx \, E} \lesssim 1
\, .
\ee
Instead, if particles are produced with large momentum, pair-production is unsuppressed provided that the de Broglie wavelength is smaller than the length-scale of field spatial gradients ($p_\x \gtrsim \w$), which is satisfied when
\be
\label{eq:adiabatic2}
\frac{\w^2}{e \qx \, E} \lesssim 1
\, .
\ee
Since the time-scale relevant for pair-production is dictated by the same length-scales discussed above, time-variations of the oscillating electric field can be ignored as well provided that the above criteria are met. For the parameter space of Fig.~\ref{fig:reach}, Eqs.~(\ref{eq:adiabatic1}) and (\ref{eq:adiabatic2}) are easily satisfied, indicating that corrections to \eq{eq:schwinger} coming from the time- and spatial-dependence of the resonant electric fields are negligible. 

We also note that effects from Pauli-blocking or Bose-enhancement in \eq{eq:schwinger} are subdominant.  After being accelerated in the cavity's electric field for a time $t_\text{acc} \sim p_\x / e \qx E \sim \sqrt{2 \pi/e q_\chi E}$, the momentum gained by mCPs is larger than that at production.  The probability of producing a particle whose wave packet overlaps within a previously produced particle's wavepacket is then $\sim P_\x \, t_\text{acc} \, d_B^3 \sim c_\x/(2 \pi) \lesssim 1$. Furthermore, for the parameter space shown in Fig.~\ref{fig:reach}, the timescale $t_\text{acc}$ is much shorter than the oscillation period of a typical SRF cavity, i.e., $t_\text{acc} \ll \w^{-1}$ for $\w \sim \GHz$ and $E \sim 50 \text{ MV} \text{ m}^{-1}$. Hence, mCPS are quickly accelerated along the direction of the electric field before escaping the cavity.

We are now in a position to calculate the production of mCPs in a driven cavity.  We will first present a toy model calculation that ignores many effects in order to illustrate the overall scaling behavior of the signal.  A more complete calculation is presented in the appendices which shows that the final results are unchanged up to $\sim \order{1}$ factors.  To begin, we take the emitter cavity to be a cylinder of length $L$ and radius $R$ that is driven in its TM$_{010}$ mode, so that the electric field points along the axis of symmetry (the $\zhat$ axis).  The resonant frequency is $\omega = \alpha_{01}/R$ where $\alpha_{01}$ is the first zero of the Bessel function $J_0$. The form of the driven fields is given by
\be
\label{eq:TMmode}
\E_\text{em} = E_\text{em} \, J_0 (\w r) \, \sin{\w t} ~ \zhat
~~,~~ 
\B_\text{em} = E_\text{em} \, J_1 (\w r) \, \cos{\w t} ~ \phihat
~,
\ee
where $r$ is the radial cylindrical coordinate.  

In our toy model estimate, up to $\order{1}$ factors, the current density of relativistic mCPs produced by the driven cavity is roughly
\be
\label{eq:jxtoy}
\jx(r,t) \sim e \qx \, L \, P_\x ~ \text{sign} (\E_\text{em} (t) \cdot \zhat) ~ \zhat
\sim 
c_\x \left( \frac{e \qx}{2 \pi} \right)^3 ~ \frac{|\E_\text{em}(r,t)|^2}{\w} ~  e^{- \pi E_\text{cr}/|\E_\text{em} (r, t)|} \, \text{sign} (\E_\text{em} (t) \cdot \zhat) ~ \zhat
~,
\ee
where $P_\x$ follows from \eq{eq:schwinger} and we have taken the length of the cavity as $L \sim 1/\w$ for reasons explained below.
The factor of $\text{sign}(E_\text{em})$ accounts for the fact that depending on the sign of the electric field, particles or antiparticles will be emitted from a fixed end of the cavity. We will approximate the above expression by taking the time-dependence of $\jx$ to be of the form
\be
|\E_\text{em}(r,t)|^2 \,  e^{- \pi E_\text{cr}/|\E_\text{em} (r, t)|} \, \text{sign} (\E_\text{em} (t) \cdot \zhat) \sim E_\text{em}^2 \, J_0 (\w r)^2 \, e^{- \pi E_\text{cr} / E_\text{em} } \, \sin{\w t}
~,
\ee
which captures the time-dependence when $E_\text{em} \gtrsim E_\text{cr}$, since $\text{sign} (\sin{\w t}) \, \sin^2{\w t}$ has an $\order{1}$ Fourier overlap with $\sin{\w t}$. Hence, our approximate form for the mCP current density is
\be
\label{eq:japprox}
\jx (r,t) \sim 
c_\x \left( \frac{e \qx}{2 \pi} \right)^3 ~ \frac{E_\text{em}^2}{\w} ~ J_0 (\w r)^2 \, e^{- \pi E_\text{cr} / E_\text{em} } \, \sin{\w t} ~ \zhat
~.
\ee

The effects that our toy model neglects are as follows.  Schwinger pair-production occurs constantly in the emitter cavity (provided that $E_\text{em} \gtrsim E_\text{cr}$), but if the magnetic fields are large compared to the electric fields, then pair-production is suppressed\footnote{In the presence of a magnetic field, the energy of a particle is increased by the energy of its lowest lying Landau level, $\w_L$.  In this case, the right-hand side of \eq{eq:SchParam1} should have $\omega_L$ added to the rest mass, $m_\x$.  For a spin-zero particle, $\omega_L \sim e \qx B/m_\x$ so that $d \rightarrow d + B/(m_\x E)$ and the expression in \eq{eq:schwinger} has an additional exponential suppression of $\text{exp}(-\pi B/E)$. For spin-1/2 (anti)particles (anti-)aligned with an external magnetic field, the first Landau level has zero energy (up to corrections from the anomalous magnetic moment), such that \eq{eq:SchParam1} is unmodified to leading order in $\order{\alpha_\text{em} \qx^2}$.} for scalar mCPs~\cite{Kim:2003qp,Cho:2003jb,Kim:2007pm} 
and, furthermore, mCPs of any spin will be deflected away from the receiver cavity. These effects are mitigated due to the fact that the resonant electric and magnetic fields are maximally out of phase (see \eq{eq:TMmode}). If the emitter cavity is long ($L \gg R$), then the total current density is a sum over mCPs that were produced at different times corresponding to distinct phases in the time-evolution of $\E_\text{em} (t)$. In our toy model, we ignore these effects. Additionally, if $L \gg R$, then the electric and magnetic fields significantly evolve in the time it takes for a typical mCP to traverse an $\order{1}$ fraction of the cavity's length, drastically modifying its trajectory.  The toy model takes this into account by approximating $L \sim \w^{-1}$. A more realistic and detailed derivation of $\jx$ is presented in Appendix~\ref{sec:milliderivation}.

\section{Response of a Receiver Cavity}
\label{sec:formalism}

We now briefly review how the shielded receiver cavity responds to the oscillating current density of mCPs.  The electric field of the receiver cavity can be decomposed as
\be
\label{eq:decomp}
\E (\xv, t) = c_n(t) \, \E_n (\xv) 
~,
\ee
where  $\E_n$ are the resonant modes of the cavity (labelled by $n$) and a sum over $n$ is implied. The cavity modes, $\E_n$, satisfy Maxwell's equations $\lap^2 \E_n = - \w_n^2 \, \E_n$, subject to the standard boundary conditions, where $\w_n$ is the resonant frequency of the $n^\text{th}$ mode. The modes also satisfy the standard orthonormality constraints 
\begin{align}
&\int_\text{rec} d^3 \xv ~ \E_n^* \cdot \E_m = \delta_{nm} \int_\text{rec} d^3 \xv ~ |\E_n|^2
~,
\end{align}
where the spatial integral is evaluated over the receiver cavity.  Taking the current to be of the form $\jx(r,t) = \jx (r) e^{i \omega t}$ and using Maxwell's equations, the expansion coefficients, $c_n$, are found to satisfy the differential equation 
\be
\label{eq:coefficients}
\ddot{c}_n + \frac{\w_n}{Q} \, \dot{c}_n  + \w_n^2 \, c_n = - i \, \w ~ e^{i \w t} ~ \frac{\int_\text{rec} d^3 \xv ~ \E_n^* \cdot \jx (r)}{\int_\text{rec} d^3 \xv ~ |\E_n|^2}
~,
\ee
where the dots denote time-derivatives and on the left-hand side we have included a dissipative energy loss term, as quantified by the large quality factor\footnote{We have assumed that energy loss from Schwinger pair-production itself is negligible compared to standard processes. This is a good approximation for the parameter space shown in Fig.~\ref{fig:reach}~\cite{Gies:2006hv}.} of the cavity, $Q$. To date, quality factors as large as $Q \sim \text{few} \times 10^{11}$ have been achieved in SRF cavities~\cite{Romanenko:2014yaa,Romanenko:2018nut}.

\eq{eq:coefficients} is simply the equation of motion of a damped harmonic oscillator that is driven by a source term as dictated by the mCP current density.  Hence, if $\w \simeq \w_n$, $\jx$ can resonantly excite power in the $n^\text{th}$ normal mode of the receiver cavity, showing that it is advantageous to tune the emitting and receiving cavities to have the same frequency. The amplitude of the excited fields in the receiver cavity is enhanced by the large quality factor and is parametrically of the form $E, B \sim Q \, j_\x / \w$. For concreteness, we take the emitter and receiver cavities to be right cylinders of equal dimensions and focus on the TM$_{010}$ mode in both cavities, as shown in \eq{eq:TMmode} (we drop the subscript $n$ below). In this case, if the emitter cavity is driven with amplitude $E_\text{em}$, then the electric field excited in the receiver cavity is determined by \eq{eq:coefficients} to be
\be
\E_\text{sig} (r, t) \simeq  E_\x ~ J_0(\w r) ~ i e^{i \w t} ~ \zhat
~,
\ee
where the toy model estimate for the amplitude, $E_\x$, is
\be
\label{eq:Ex}
E_\x \sim c_\x \left( \frac{e \qx}{2 \pi} \right)^3 ~ \frac{Q \, E_\text{em}^2}{\w^2} ~ e^{- \pi E_\text{cr} / E_\text{em}} ~ \eta_j
~.
\ee
Above, $\eta_j$ is an $\order{1}$ mode-dependent factor given by
\be
\label{eq:etaj1}
\eta_j = \frac{\int_0^R dr ~ r ~ J_0 (\w r)^3}{\int_0^R dr ~ r ~ J_0 (\w r)^2} \simeq 0.72
\, .
\ee
The total signal power in the receiver cavity is
\be
P_\text{sig} \simeq \frac{\w}{Q} \int_\text{rec} d^3 \xv ~ |\E_\text{sig}|^2 \sim  \frac{\w}{Q} \, \eta_{_V} \, E_\x^2 \, V_\text{cav} 
~,
\ee
where $V_\text{cav}$ is the total volume of the receiver cavity and $\eta_{_V}$ is an additional $\order{1}$ mode-dependent factor given by
\be
\label{eq:etav1}
\eta_{_V} = \frac{2}{R^2} \int_0^R dr ~ r ~ J_0(\w r)^2 \simeq 0.27
\, .
\ee
An analogous version of this calculation utilizing a more complete estimate for $\jx$ is presented in Appendix~\ref{sec:cavityderivation}.

\section{Sensitivity}
\label{sec:reach}

The signal-to-noise ratio scales as $\text{SNR} = P_\text{sig} / P_\text{noise} \propto \qx^6$ where $P_\text{noise}$ is the total noise power. In calculating the reach of a future cavity experiment, we assume that noise is controlled by thermal fluctuations of the receiver cavity, which is expected to dominate over the intrinsic noise of the readout device, such as a SQUID magnetometer~\cite{Graham:2014sha,fnalth,fnalex}. In this case, the noise power is approximately  $P_\text{noise} \simeq T / t_\text{int}$, where $T$ is the temperature of the receiver cavity and $t_\text{int}$ is the total integration time of the experiment. The fact that the SNR increases linearly with integration time (as opposed to $\text{SNR} \propto \sqrt{t_\text{int}}$, as is the case for, e.g., resonant axion searches for a background dark matter field) is due to the fact that that one can measure the frequency and phase of the emitter cavity which in turn allows for a determination of the signal phase in the receiver cavity. As discussed in Ref.~\cite{Graham:2014sha}, this allows for a measurement of the signal field (as opposed to power) which grows as $\sqrt{t_\text{int}}$ and thus a signal power that scales linearly in total integration time. Without a measurement of the emitter phase, the noise power is instead $P_\text{noise} \sim T \sqrt{\w / (Q t_\text{int})}$, which results in noise levels roughly $\sim 100$ times larger for the experimental parameters adopted in this work. Here, we assume that the emitter phase has been measured. However, since $P_\text{sig} \propto \qx^6$, not doing so only leads to a factor of $\sim 2$ decrease in the projected reach of Fig.~\ref{fig:reach}.

Assuming that the electric field of the driven emitter cavity is much larger than the critical field strength for Schwinger pair-production of mCPs ($E_\text{em} \gg E_\text{cr}$), a signal-to-noise ratio of $\text{SNR} \gtrsim 1$ is equivalent to
\be
\label{eq:reach}
\qx \gtrsim \order{10^{-13}} ~ c_\x^{-1/3} \, \left( \frac{E_\text{em}}{50 \text{ MV}/ \text{m}} \right)^{-2/3} \left( \frac{Q}{10^{12}} \right)^{-1/6} \left( \frac{\w}{\GHz} \right)^{1/2} \left( \frac{V_\text{cav}}{\text{m}^3} \right)^{-1/6} \left( \frac{T}{10 \text{ mK}} \right)^{1/6} \left( \frac{t_\text{int}}{\text{year}} \right)^{-1/6}
~.
\ee
To date, the most stringent existing laboratory based constraints are derived from searches for vacuum magnetic birefringence by the PVLAS experiment~\cite{Ahlers:2006iz,DellaValle:2014xoa}. The projected reach of Eq.~(\ref{eq:reach}) is sensitive to couplings $\sim 10^6$ smaller than those currently excluded by PVLAS. In estimating the reach of SRF cavities, we have assumed that the driving and resonant frequencies of the emitter and receiver cavities are degenerate. Controlling this degeneracy for the extremely narrow resonances of SRF cavities is one of the main experimental feats that the FNAL setup is expected to overcome. In particular, mechanical vibrations of the cavity from seismic noise and vibrations from the liquid helium cyrogenic system can lead to small time-dependent variations of the cavity's resonant frequencies. Mitigating this mode-wobbling to one part in $Q$ necessitates controlling the positions of the cavity walls to sub-nanometer precision~\cite{fnalex}. 

\section{Discussion and Conclusions}
\label{sec:conclusion}

In this work, we have shown that light-shining-through-walls experiments can have world-leading sensitivity to light millicharged particles that are Schwinger pair-produced due to the large electric fields of superconducting radio-frequency cavities. A qualitatively similar discussion to the one illustrated here was presented in Refs.~\cite{Gies:2006hv,Jaeckel:2007hi}. These previous studies considered cavities of smaller geometric size and weaker field gradients. In starker contrast is the fact that Ref.~\cite{Jaeckel:2007hi} quantified detectable signal levels as total millicharge currents greater than $\sim \text{nA} - \mu \text{A}$. Adopting this criteria for quantifying detectability, we agree with Refs.~\cite{Gies:2006hv,Jaeckel:2007hi} which found that this corresponds to $\qx \gtrsim 10^{-7} - 10^{-6}$, respectively. In particular, the parametric expression in \eq{eq:jxtoy} implies that
\be
\label{eq:them}
j_\x R^2 \sim \mu \text{A} \times c_\x \left( \frac{\qx}{10^{-6}} \right)^{3} \left( \frac{E_\text{em}}{15 \text{ MV} \text{ m}^{-1}} \right)^{2} \left( \frac{\w}{\GHz} \right)^{-1} \left( \frac{R}{10 \ \cm} \right)^{2} 
~,
\ee
which is normalized to the cavity parameters in Ref.~\cite{Jaeckel:2007hi}. In this study, we have provided a detailed estimate, which shows that a thermal-noise limited superconducting radio-frequency setup will in fact be sensitive to much smaller couplings. The generalized version of the toy model calculation of Sec.~\ref{sec:formalism} shows that near-future experimental setups will attain sensitivity to oscillating currents as small as
\be
\label{eq:us}
j_\x R^2 \sim 10^{-24} \text{ A} \times \left( \frac{Q}{10^{12}} \right)^{-1/2} \left( \frac{\w}{\GHz} \right)^{1/2} \left( \frac{V_\text{cav}}{500 \ \cm^3} \right)^{-1/2} \left( \frac{T}{ 10 \text{ mK}} \right)^{1/2} \left( \frac{t_\text{int}}{\text{year}} \right)^{-1/2} \left( \frac{R}{10 \ \cm} \right)^2
~.
\ee
The many orders of magnitude in difference between Eqs.~(\ref{eq:them}) and (\ref{eq:us}) serves to partially explain why the projections shown in Fig.~\ref{fig:reach} are enhanced by a factor of $\sim 10^{18/3} \sim 10^6$ compared to the estimate in Refs.~\cite{Gies:2006hv,Jaeckel:2007hi}. 

It is important to note that while terrestrial experiments utilizing superconducting radio-frequency cavities could provide the best laboratory constraints on new electrically charged particles, very powerful limits have also been derived from the consideration of various astrophysical and cosmological processes, such as stellar cooling, SN1987A, Big Bang nucleosynthesis, and the cosmic microwave background (see Ref.~\cite{Vogel:2013raa} and references therein). In Appendix~\ref{sec:cavityderivation}, we discuss models that significantly alleviate these bounds, opening up the parameter space shown in Fig.~\ref{fig:reach}. We stress that independent of such model building, it is sufficiently motivating that near-future light-shining-through-walls experiments could provide the best  terrestrial constraints on light millicharged particles without modifying their planned geometry or data acquisition.

It is very exciting that a light-shining-through-walls experiment utilizing superconducting radio-frequency cavities is currently under construction at FNAL. The ability to search for ultralight dark photons has been the main physics motivation for such a setup up to this point~\cite{Graham:2014sha,fnalth}.  In fact, an emitter-receiver cavity geometry identical to that shown in Fig.~\ref{fig:cartoon} is also optimal for detecting the enhanced longitudinal mode of dark photons much lighter than $\w \sim \GHz$.  Importantly, how this signal scales with various experimental parameters differs compared to the millicharge-induced one discussed in this work. In particular, while the electric field signal in the receiver cavity scales as $E_\text{sig} \propto E_\text{em}^2$ for millicharges in \eq{eq:Ex}, for light dark photons this is modified to $E_\text{sig} \propto E_\text{em}$~\cite{Graham:2014sha}. 
Hence, in the exciting event of an observed signal, various new physics explanations could be differentiated by slightly varying the power that is driven into the emitter cavity.

\section*{Acknowledgements}
We would like to thank Aaron Chou, Roni Harnik, and Junwu Huang for valuable conversations. AB is supported by the James Arthur Fellowship. AH is supported in part by the NSF under Grant No. PHY-1914480 and by the Maryland Center for Fundamental Physics (MCFP).

\appendix

\section{Derivation of Millicurrents from Schwinger Pair-Production}
\label{sec:milliderivation}

In this appendix, we calculate the charge and current density of pair-produced mCPs, $\x^\pm$. The starting point of this calculation assumes that after production, the mCPs follow localized trajectories in spacetime. This is equivalent to the classical limit of a continuous fluid. Later, in Appendix~\ref{sec:mCPs}, we comment on the generalization of this calculation when this assumption is not valid within the context of quantum mechanical fluids. 

The number of particles of charge $\alpha e \qx$ ($\alpha = \pm 1$) that are produced per unit volume and unit time is denoted as
\be
P_\alpha (\xv , t) = \frac{dN_\alpha}{d^3 \xv \, dt}
~.
\ee
Therefore, the infinitesimal number density of point particles that are created at an initial position $\xv_i$ at time $t_i$ is 
\be
\label{eq:nx1}
d n_\alpha (\xv, t_i) = dN_\alpha (\xv_i , t_i) ~ \delta^3(\xv - \xv_i) = d^3 \xv_i ~ dt_i ~ P_\alpha (\xv_i , t_i) ~ \delta^3(\xv - \xv_i)
~.
\ee
In order to time-evolve this number density to later times ($t > t_i$), in \eq{eq:nx1} we promote $\xv_i$ to a time-dependent coordinate, i.e.,
\be
\delta^3(\xv - \xv_i) \to \delta^3 (\xv - \xv_\alpha (\xv_i, t_i, t))
\ee
where $\xv_\alpha (\xv_i, t_i, t)$ is the trajectory of a particle of charge $\alpha e \qx$ at time $t$, given that is was produced at position $\xv_i$ and time $t_i$. Time-evolving \eq{eq:nx1} in this manner, we have
\be
d n_\alpha (\xv, t) =d^3 \xv_i ~ dt_i ~ P_\alpha (\xv_i , t_i) ~ \delta^3 (\xv - \xv_\alpha (\xv_i, t_i, t))
~.
\ee
The total number density, $n_\x$, is then obtained by summing over species of either charge ($\alpha = \pm 1$) and all possible initial positions ($\xv_i$) and times ($t_i < t$),
\be
\label{eq:nx2}
n_\x (\xv , t) = \sum\limits_{\alpha = \pm 1}\int d^3 \xv_i ~ dt_i ~ P_\alpha (\xv_i , t_i) ~ \delta^3 (\xv - \xv_\alpha (\xv_i, t_i, t))
~.
\ee
In general, the $\xv_i$ integral is to be performed over all of space. However, for particles that are pair-produced from the large electric fields of resonant cavities, the spatial integral only has significant weight over the interior of the cavity itself since $P_\alpha$ is exponentially suppressed elsewhere. In order to take into account that $\x^\pm$ are pair-produced with a spread of initial velocities ($\vv_i$) and that their trajectories ($\xv_\alpha$) depend explicitly on $\vv_i$, we also add to \eq{eq:nx2} a  weighted sum over $\vv_i$,
\be
\label{eq:nx3}
n_\x (\xv , t) = \sum\limits_{\alpha = \pm 1} \int d^3 \xv_i ~ dt_i ~ d^3 \vv_i ~ f_\alpha (\vv_i ; \xv_i , t_i) ~ P_\alpha (\xv_i , t_i) ~ \delta^3 (\xv - \xv_\alpha (\xv_i, \vv_i, t_i, t))
~,
\ee
where $f_\alpha (\vv_i ; \xv_i , t_i)$ is the unit-normalized distribution of initial velocities. The total current density, $\jx$, is derived in a nearly identical manner. Compared to \eq{eq:nx3}, the only difference is that the sum over $\alpha$ inherits a factor of charge ($\alpha e \qx$) and the integrand involves an overall factor of the time-evolved velocity, $\vv_\alpha (\xv_i, \vv_i, t_i, t)$,
\begin{align}
\label{eq:JxExact}
\jx (\xv , t) = \sum\limits_{\alpha = \pm 1} \alpha \, e \qx \int d^3 \xv_i ~ dt_i ~ d^3 \vv_i ~ f_\alpha (\vv_i ; \xv_i , t_i) ~   \vv_\alpha (\xv_i , \vv_i, t_i,  t) ~ P_\alpha (\xv_i , t_i) ~ \delta^3 (\xv - \xv_\alpha (\xv_i, t_i, \vv_i, t))
~.
\end{align}

The above discussion is accurate in the limit that the number density, $n_\x$, is large and the produced $\x^\pm $ population can be approximated as a continuous collisionless fluid. However, \eq{eq:JxExact} in its current form is not immediately useful in providing an analytic handle on $\jx$. As discussed in Sec.~\ref{sec:schwinger}, in most regions of parameter space in Fig.~\ref{fig:reach}, $\x^\pm$ pairs are produced and accelerated to ultra-relativistic speeds along the direction of the electric field. For the TM$_{010}$ resonant modes of \eq{eq:TMmode}, the electric field of the driven cavity is purely in the longitudinal ($\zhat$) direction and is $z$-independent. Hence, in order to simplify \eq{eq:JxExact}, we approximate the initial velocity, $\vv_i$, and the time-evolved trajectories, $\xv_\alpha$, as purely relativistic along the $\zhat$ direction, i.e.,
\begin{align}
\label{eq:ultraapprox}
f_\alpha (\vv_i ; \xv_i , t_i)  &\simeq \delta^3 \Big(\vv_i - \alpha ~ \text{sign} \big(\E_\text{em} (\xv_i,t_i) \cdot \zhat \big) \Big) 
\nl
\xv_\alpha (\xv_i, t_i, \vv_i, t) &\simeq \xv_i + \alpha ~ \text{sign} \big(\E_\text{em} (\xv_i,t_i) \cdot \zhat \big) ~ (t - t_i) ~ \zhat
~.
\end{align}
Decomposing $\xv = (\xv_\perp, z)$ and using the delta-functions in Eqs.~(\ref{eq:JxExact}) and (\ref{eq:ultraapprox})  to perform the integrals over $\vv_i$, $\xv_{\perp i}$, and $t_i$, we find
\begin{align}
\label{eq:JxApprox}
\jx (\xv_\perp , z , t) \simeq \ e \qx \, \zhat \int_0^L dz_i ~ P \big(\xv_\perp, z_i , t - (z - z_i) \big) ~ \text{sign} \Big( \E_\text{em} \big( \xv_\perp,  t - (z - z_i)\big) \cdot \zhat \Big)
~,
\end{align}
where the length of the cylindrical cavity is taken to run from $z=0$ to $z=L$. Note that within the integral over $z_i$ in \eq{eq:JxApprox}, the production rate and the electric field of the emitter cavity are evaluated at the retarded time $t - (z-z_i)$. Since Schwinger pair-production is independent of the sign of the millicharge, we have also dropped the $\alpha$ subscript for the production rate, $P$. 

The rate for Schwinger pair-production scales as the electric field squared, $P \propto |\E_\text{em}|^2$. In order to analytically simplify \eq{eq:JxApprox}, we parametrize the Schwinger production rate as 
\be
P (\xv , t) = \frac{c_\x}{(2 \pi)^3}  ~ (e \qx)^2 ~ |\E_\text{em}(\xv, t)|^2 ~ e^{- \pi E_\text{cr} / |\E_\text{em} (\xv, t)|}
~,
\ee
where $c_\x = 1 \ (1/2)$ for fermionic (scalar) $\x$ and $E_\text{cr}$ is the critical electric field strength. For an oscillating TM$_{010}$ electric field of the form
\be
\E_\text{em} (\xv, t) = E_\text{em} (r) \, \sin{ \w t} ~ \zhat
~,
\ee
we then approximate the integrand of \eq{eq:JxApprox} using
\be
\label{eq:intapprox}
P (\xv, t) ~ \text{sign}\big(\E_\text{em} (\xv, t) \cdot \zhat \big) \simeq \frac{c_\x}{(2 \pi)^3} ~ (e \qx)^2 ~ E_\text{em}(r)^2 ~ e^{- \pi E_\text{cr} / E_\text{em}(r)}
~ \sin{\w t}
~,
\ee
which accurately captures the full time-dependence, as discussed in Sec.~\ref{sec:schwinger}. Using \eq{eq:intapprox} in \eq{eq:JxApprox}, we then find
\be
\label{eq:JxApprox2}
\jx (\xv, t) \simeq \frac{c_\x}{4 \pi^3} ~ (e \qx)^3 \, E_\text{em}(r)^2 ~ \frac{\sin{\varphi}}{\w} 
~ e^{- \pi E_\text{cr} / E_\text{em}(r)} 
~ \sin{(\w (t-z) + \varphi)} ~ \zhat
~,
\ee
where the overall phase is given by $\varphi = \w L/2$.

We have checked that in the absence of magnetic fields, \eq{eq:JxApprox2} is an accurate approximation to \eq{eq:JxExact}. However, the presence of magnetic fields can significantly alter the pair-production rate as well as the trajectories of mCPs before escaping the emitter cavity. In particular, Schwinger pair-production of spin-0 particles is exponentially suppressed when $B_\text{em} \gg E_\text{em}$~\cite{Kim:2003qp,Kim:2007pm}. Furthermore, even for particles that are efficiently pair-produced when $E_\text{em} \gg B_\text{em}$, they may encounter field-configurations for which $B_\text{em} \gg E_\text{em}$ before they escape the cavity. 
If pair-produced particles encounter fields such that $B_\text{em} \gg E_\text{em}$ and if their gyroradius is much smaller than the geometric size of the cavity, then their velocities typically develop significant radial components before exiting the cavity. For such trajectories, the radial diffusion of $\x^\pm$ significantly reduces the size of the millicurrent, $j_\x$, near the downstream receiver cavity. 

These effects are mitigated by the fact that Maxwell's equations implies that electric and magnetic fields are maximally out of phase in a resonant cavity. Hence, maximal pair-production occurs when magnetic fields are at a temporal minimum. 
Taking full numerical account of the magnetic fields is beyond the scope of this work.  Instead, we numerically evaluate the current density, assuming that mCPs negligibly contribute to $\jx$ if they encounter regions in which $B_\text{em} (t) \gtrsim E_\text{em} (t)$ at any moment before escaping the emitter cavity. We then find that for a cylindrical cavity of similar radius and length ($R \simeq L$) that is driven in the TM$_{010}$ mode, \eq{eq:JxExact} is well-approximated numerically if the expression of \eq{eq:JxApprox2} is restricted to the radial region $r \lesssim R/2$.  Hence, we take
\be
\label{eq:JxApprox3}
\jx (\xv, t) \simeq \frac{c_\x}{4 \pi^3} ~ (e \qx)^3 \, E_\text{em}(r)^2 ~ \frac{\sin{\varphi}}{\w} 
~ e^{- \pi E_\text{cr} / E_\text{em}(r)} ~ \Theta (R/2-r)
~ \sin{(\w (t-z) + \varphi)} ~ \zhat
~,
\ee
where $\Theta$ is the Heaviside step-function.  This expression for $j_\x$, which approximately accounts for the requirement that $E_\text{em} \gtrsim B_\text{em}$ before an mCP escapes the emitter cavity, is used to calculate the projected sensitivities shown in Fig.~\ref{fig:cartoon}.

\section{Cavity Response}
\label{sec:cavityderivation}

In this appendix, we derive the response of the receiver cavity to the oscillating mCP current. The calculation is nearly identical to that shown in Sec.~\ref{sec:formalism}, except that instead of using the toy model expression for $\jx$ in \eq{eq:japprox}, we use \eq{eq:JxApprox3}.

In this case, if the emitter cavity is driven with amplitude $E_\text{em}$, then the electric field excited in a receiver cavity placed a distance $d$ from the front-end of the emitter cavity  is approximately
\be
\E_\text{sig} (\xv, t) \simeq  E_\x ~ J_0(\w r) ~ i \, e^{i (\w (t-d) - \varphi)} ~ \zhat
~.
\ee
where the amplitude of the field is given by
\be
E_\x \simeq \frac{\eta_j \, Q}{2 \pi^3} ~ c_\x \, (e \qx)^3 ~ E_\text{em}^2 ~ e^{- \pi E_\text{cr} / E_\text{em}} ~ \frac{\sin^2{\varphi}}{\w^3 L}
~.
\ee
$\eta_j$ is a mode-dependent $\order{1}$ factor. For the TM$_{010}$ mode, as considered here, it is given by
\be
\eta_j = \frac{\int_0^{R/2} dr ~ r ~ J_0(\w r)^3}{\int_0^R dr ~ r ~ J_0(\w r)^2} \simeq 0.55
~.
\ee
Note that the region of integration in this expression for $\eta_j$ is slightly modified compared to that of \eq{eq:etaj1}, due to the inclusion of the Heaviside step-function in \eq{eq:JxApprox3}. The signal power is then approximately
\begin{align}
P_\text{sig} &\simeq \frac{\w}{Q} \int_\text{em} d^3 \xv ~ |\E_\text{sig}|^2 \simeq \frac{\w}{Q} ~ \eta_{_V} \, E_\x^2 ~ V_\text{cav} 
~,
\end{align}
where $\eta_{_V}$ is the same as in \eq{eq:etav1}. The rest of the calculation is identical to Sec.~\ref{sec:formalism}, which ultimately leads to a projected reach that is nearly identical to \eq{eq:reach} up to $\lesssim \order{1}$ factors.

\section{Alleviating Astrophysical and Cosmological Bounds}
\label{sec:stellar}

Powerful limits on mCPs have been derived from various astrophysical and cosmological processes~\cite{Vogel:2013raa}. The most stringent of these come from considerations of stellar cooling, which is sensitive to millicharges $\qx \gtrsim \text{few} \times 10^{-14}$ for the simplest of such models. In this appendix, we briefly discuss a specific model of mCPs presented in Ref.~\cite{Masso:2006gc} that alleviates such constraints. As mentioned in Sec.~\ref{sec:intro}, small millicharge couplings can naturally emerge from the kinetic mixing of a light dark photon. In this section, we instead focus on a model that involves two dark photons, $\Ap_1$ and $\Ap_2$ of mass $m_1$ and $m_2$, and a vector-like pair of fermions $\x$ and $\x^c$ with charges $(1,-1)$ and $(-1,1)$ under each dark photon, respectively. For concreteness, we assume that $m_2 = 0$, although our conclusions are qualitatively unchanged for $m_2 \lesssim \text{meter}^{-1} \sim 10^{-7} \ \eV \ll m_1$.
This theory therefore has a $\mathbb{Z}_2$ symmetry under which $\Ap_1 \Leftrightarrow \Ap_2$ and $\x \Leftrightarrow \x^c$, which is  softly broken by $m_1 \ne 0$.  Since the breaking is soft, any correction to the symmetry will be suppressed by the dimension-two mass-squared parameter.

Denoting the SM photon as $A_\gamma$, the most general Lagrangian allowed by these symmetries is
\be
\label{eq:KinMix1}
\mathscr{L} = -\frac{1}{4} \boldsymbol{F}^T_{\mu \nu}
\begin{pmatrix} 
1 & \eps & \eps \\
\eps & 1 & 0\\
\eps & 0 & 1
\end{pmatrix} 
\boldsymbol{F}_{\mu \nu} + \frac{1}{2} \boldsymbol{A}^T_\mu 
\begin{pmatrix} 
m_\gamma^2 & 0 & 0 \\
0 & m_1^2 & 0\\
0 & 0 & 0
\end{pmatrix}
\boldsymbol{A}_\mu + e J_{em}^\mu A_\gamma^\mu + e^\prime J_1^\mu A_1^{\prime \mu} + e^\prime J_2^\mu A_2^{\prime \mu} ,
\ee
where we have used the notation 
$ \boldsymbol{A}^\mu = 
\begin{pmatrix} 
A_\gamma^\mu & A_1^{\prime \mu} & A_2^{\prime \mu}
\end{pmatrix}^T$ 
for the gauge fields and 
$ \boldsymbol{F}^{\mu \nu} = 
\begin{pmatrix} 
F_\gamma^{\mu \nu} & F_1^{\prime \mu \nu} & F_2^{\prime \mu \nu}
\end{pmatrix}^T$
for the corresponding field-strengths.\footnote{In the Lagrangian above, we have assumed that any small amount of kinetic mixing between $\Ap_1$ and $\Ap_2$, denoted as $\tilde{\eps}$, is diagonalized away by shifting the massless $\Ap_2$ field, which then modifies the $\Ap_2$ interaction term in \eq{eq:KinMix1}. Ignoring this effect amounts to dropping $\order{\eps \, \tilde{\eps}}$ terms in \eq{eq:qeff} below, which we assume are subdominant.}  $J_\text{em}^\mu$ is the SM electromagnetic current density, while $J_{1,2}^\mu$ corresponds to $A_{1,2}^{\prime}$. For the charge assignment above, in four-component notation they are given by $J_1^\mu = \bar{\x} \gamma^\mu \x$ and $J_2^\mu = - \bar{\x} \gamma^\mu \x$. In \eq{eq:KinMix1}, the presence of the mass parameter, $m_\gamma$,  is a stand-in for either the plasma or Debye mass of the SM photon, depending on the process of interest~\cite{Knapen:2017xzo,Dvorkin:2019zdi}. The kinetic and mass terms can be diagonalized after replacing $\boldsymbol{A}^\mu \to U \boldsymbol{A}^\mu$, where
\be
U = \begin{pmatrix} 
1 & \eps \frac{m_1^2}{m_\gamma^2 - m_1^2} & 0 \\
\eps \frac{m_\gamma^2}{m_1^2-m_\gamma^2} & 1 & 0\\
-\eps & 0 & 1
\end{pmatrix}
~,
\ee
to leading order in $\eps$.
From this, we can see that the effective charge of $\x$ under the SM photon is
\be
\label{eq:qeff}
\qx \simeq \frac{\eps \, e^\prime}{e} \left( \frac{m_1^2}{m_1^2 - m_\gamma^2} \right)
\simeq \frac{\eps \, e^\prime}{e} \times \begin{cases} 1  & \text{when } m_\gamma \ll m_1 \text{ (in vacuum)} \\ (m_1 / m_\gamma)^2  & \text{when } m_1 \ll m_\gamma \text{ (in plasma)} ~, \end{cases}
\ee
due to the partial cancellation of the photon component of $\Ap_1 - \Ap_2$, the same linear combination of fields to which $\x$ couples. Therefore, since a laboratory is approximately void of charged particles, the plasma/Debye mass is negligible ($m_\gamma \ll m_1$) and the effective millicharge of $\x$ depends on the kinetic mixing and hidden sector gauge coupling as expected. On the other hand, in a dense plasma, such as the interior of a star, supernova, or in the early universe, we can choose $m_1 \ll m_\gamma$ such that $\qx$ is very suppressed, alleviating the constraints derived from the considerations of such systems. In this case, if additionally $m_1 \gg \text{meter}^{-1} \sim 10^{-7} \ \eV$, then on laboratory length-scales only $\Ap_2$ is long-ranged and $\qx$ is not screened. In this manner, astrophysical and cosmological constraints are weakened by a factor of $(m_1/m_\gamma)^2$, while laboratory based experiments are unaffected.  

A critical feature of this mechanism that allows for the cancellation in \eq{eq:qeff} is the fact that the kinetic mixing parameter, $\eps$, and the hidden sector gauge couplings, $e^\prime$, are the same between the two hidden sectors.
The soft $\mathbb{Z}_2$-breaking nature of $m_1$ implies that any corrections to these relations must vanish as $m_1 \rightarrow 0$.
For example, if $m_1 > m_\x,  q$ (where $q \sim \rm{keV}$ is the energy scale associated with stellar constraints), then there is additional RG running for the $A_2$ gauge field coupling so that $\Delta \alpha_D \sim \alpha_D^2 \log (m_1^2/\text{max} (m_\x^2 , q^2))$.  
The region of parameter space that suffers the least from the soft symmetry breaking is when $q > m_\x, m_1$.  By dimensional analysis, the corrections scale as $\Delta \alpha_D \sim \alpha_D \, m_1^2/q^2$ or smaller, and the $\mathbb{Z}_2$ symmetry breaking can be easily suppressed to sufficient levels.

\section{Quantum Mechanical Description}
\label{sec:mCPs}

An important and implicit part of the derivation of the response of the receiver cavity is that there was never a measurement of the number of emitted mCPs.  The reason can be seen by observing the average number of mCPs per volume of the cavity
\be
\label{eq:nx}
N_\x \sim \frac{(e \qx)^2 c_\x}{(2 \pi)^3} ~ \frac{E_\text{em}^2}{\w} ~ V_\text{cav}  \sim \order{0.1} \times c_\x \left( \frac{\qx}{10^{-12}} \right)^{2} \left( \frac{E_\text{em}}{50 \text{ MV} \text{ m}^{-1}} \right)^{2} \left( \frac{\w}{\text{GHz}} \right)^{-1} \left( \frac{V_\text{cav}}{10^{-3} \text{ m}^{3}} \right)
~.
\ee
From this, one can see that over a large swath of the parameter space shown in Fig.~\ref{fig:reach}, there is on average less than one mCP within the receiver cavity volume at a fixed time. In this case, if the mCPs were projected onto a number eigenstate, then the resulting Poisson fluctuations would constitute a crucial noise source when $N_\x \lesssim \order{10}$. However, it is only when the mCPs interact that they can be projected onto a number eigenstate.  As the mCPs pass through any shielding with negligible interactions, the only opportunity to be projected onto a number state occurs when the mCPs are accelerated by the electric fields in the emitter cavity.  For example, if the mCPs completely discharge the electric field of the emitter cavity, it constitutes a ``measurement" of the number of mCPs that are produced.

As Poisson fluctuations are potentially substantial when $N_\x$ is small, the relevant question is if the source cavity can detect the production of a single mCP, since this determines whether the produced mCP state can be described in terms of number eigenstates.  Upon production, an mCP is accelerated to an energy $\sim e \qx E_\text{em} L$ after traversing a distance $L$ and hence absorbing
\be
N_\text{absorb} \sim \frac{e \qx E_\text{em} L}{\omega}
\ee
photons from the cavity.  The emitter cavity has a large electric field and is not in a photon number eigenstate.  The average number of photons in an emitter cavity of volume $V_\text{cav}$ is
\be
N_\text{cavity} \sim \frac{E_\text{em}^2 \, V_\text{cav}}{\omega}
\ .
\ee
From a quantum mechanical viewpoint, a state in which the cavity possess $N_\text{cavity}$ photons and $N_\text{cavity} + \Delta$ photons are indistinguishable (and hence have a large overlap) provided that $\Delta \lesssim \sqrt{N_\text{cavity}}\ $.
In this case, the cross terms describing the interference between such states are non-negligible and a ``measurement" has not been made.\footnote{A useful analogy is the double slit experiment, where the interference pattern vanishes if a detector can detect which slit the electron travels through. In this language, it is clear that a measurement occurs if the detector eigenstates corresponding to when the electron goes through either path are orthogonal to each other, thereby removing the interference cross term.}  Hence, the emitter cavity cannot distinguish between the number of mCPs produced if $N_\text{absorb} \lesssim \sqrt{N_\text{cavity}}$, which corresponds to
\be
\label{eq:qxmeasure1}
\qx \lesssim \frac{\sqrt{\omega V_\text{cav}}}{e L} \sim \order{1} \times \left( \frac{\w}{\text{GHz}} \right)^{1/2} \left( \frac{V_\text{cav}}{\text{m}^3} \right)^{1/2} \left( \frac{L}{\text{m}} \right)^{-1}
\, .
\ee
The upper bound in Eq.~(\ref{eq:qxmeasure1}) is consistent with the fact that the production of a single mCP cannot be inferred from the measured energy loss of the emitter cavity, as encapsulated in its quality factor, $Q$. In particular, demanding that a single mCP is a smaller energy sink than standard processes leads to 
\be
q_\x \lesssim \frac{E_\text{em} \, V_\text{cav}^{2/3}}{e \, Q} \sim \order{10^3} \times \left( \frac{E_\text{em}}{50 \text{ MV} \text{ m}^{-1}} \right) \left( \frac{V_\text{cav}}{\text{m}^3} \right)^{2/3} \left( \frac{Q}{10^{12}} \right)^{-1}
\ee
Hence, when Eq.~(\ref{eq:qxmeasure1}) holds (which is the case for the entire parameter space shown in Fig.~\ref{fig:reach}), the proper treatment is to consider the stream of pair-produced mCPs as a quantum mechanical wave (e.g., as giving rise to a fixed electromagnetic field) rather than as a discrete flow of localized point-particles. 

In this regime, the produced current of mCPs matches that of the classical particle-based result derived in Appendix~\ref{sec:milliderivation}, even if $N_\x \lesssim \order{1}$. To see this, note that the main difference between the evolution of classical and quantum phase space stems from the uncertainty principle; ``quantum pressure" resists the localization of a particle with momentum $p$ on length-scales smaller than $1/p$. For instance, this is well-known in the context of fuzzy dark matter (see, e.g., Ref.~\cite{Hui:2016ltb}), where the quantum pressure prevents localization of dark matter on scales smaller than the de Broglie wavelength.  

In the non-relativistic limit, it has been shown that the quantum phase space distribution (as dictated by the Schr{\"o}dinger equation) is the same as the classical phase space distribution (as dictated by the Hamilton-Jacobi equation) up to a contribution from the quantum potential, $U_Q \sim (\grad^2 \sqrt{n_\x}) / (m_\x \sqrt{n_\x}) \sim \w^2 / m_\x$, where $n_\x$ is the mCP number density. For non-relativistic mCPs, the classical fluid formalism outlined in Appendix~\ref{sec:milliderivation} is valid provided that $\grad U_Q \sim \w U_Q$ is subdominant compared to the classical electromagnetic force ($e \qx E_\text{em}$), which occurs when
\be
\label{eq:QM1}
m_\x \, e \qx \, E_\text{em} \gtrsim \w^3
\, .
\ee
This equation can be understood intuitively. Consider the length-scale $L_n \sim 1/\omega$ over which $n_\x$ changes by an $\order{1}$ fraction. After traversing a distance $L_n$, mCPs gain a kinetic energy of $p_\x^2/ 2 m_\x \sim e \qx  E_\text{em} L_n$ such that $p_\x \gtrsim \sqrt{ m_\x e \qx E_\text{em} L_n}$.  Requiring that $p_\x \gtrsim 1/L_n$ so that the de Broglie wavelength does not wash out particle localization within the scale of density gradients then leads directly to \eq{eq:QM1}.
This same intuition can be applied to the parameter space in Fig.~\ref{fig:reach}, most of which lies in the relativistic regime. Analogous to before, but now requiring that $p_\x \sim e \qx E_\text{em} L_n \gtrsim 1/L_n$, we obtain
\be
\qx \gtrsim \frac{\omega^2}{e E_\text{em}} \sim \order{10^{-14}} \times \left( \frac{\w}{\text{GHz}} \right)^{2} \left( \frac{E_\text{em}}{50 \text{ MV} \text{ m}^{-1}} \right)^{-1}
\, ,
\ee
which is valid for the couplings shown in Fig.~\ref{fig:reach}. Thus, we expect \eq{eq:JxApprox3} to hold over the parameter space considered in this work.

\bibliography{schwinger}
\bibliographystyle{utphys}

\end{document}